\documentclass[aps,prd,twocolumn,groupedaddress,showpacs,preprintnumbers,draft]
{revtex4}
\usepackage{epsf} 
\begin{document}
\preprint{MCGILL-22-04}
\def\Box{\nabla^2}  
\def\ie{{\em i.e.\/}}  
\def\eg{{\em e.g.\/}}  
\def\etc{{\em etc.\/}}  
\def\etal{{\em et al.\/}}  
\def\S{{\mathcal S}}  
\def\I{{\mathcal I}}  
\def\mL{{\mathcal L}}  
\def\H{{\mathcal H}}  
\def\M{{\mathcal M}}  
\def\N{{\mathcal N}} 
\def\O{{\mathcal O}} 
\def\cP{{\mathcal P}} 
\def\R{{\mathcal R}}  
\def\K{{\mathcal K}}  
\def\W{{\mathcal W}} 
\def\mM{{\mathcal M}} 
\def\mJ{{\mathcal J}} 
\def\mP{{\mathbf P}} 
\def\mT{{\mathbf T}} 
\def\mR{{\mathbf R}}
\def\mS{{\mathbf S}}
\def\mX{{\mathbf X}}
\def\mZ{{\mathbf Z}}
\def\eff{{\mathrm{eff}}}  
\def\Newton{{\mathrm{Newton}}}  
\def\bulk{{\mathrm{bulk}}}  
\def\brane{{\mathrm{brane}}}  
\def\matter{{\mathrm{matter}}}  
\def\tr{{\mathrm{tr}}}  
\def\nr{{\mathrm{normal}}}  
\def\implies{\Rightarrow}  
\def\half{{1\over2}}  
\newcommand{\da}{\dot{a}}
\newcommand{\db}{\dot{b}}
\newcommand{\dn}{\dot{n}}
\newcommand{\dda}{\ddot{a}}
\newcommand{\ddb}{\ddot{b}}
\newcommand{\ddn}{\ddot{n}}
\newcommand{\ba}{\begin{array}}
\newcommand{\ea}{\end{array}}
\def\be{\begin{equation}}
\def\ee{\end{equation}}
\def\bea{\begin{eqnarray}}
\def\eea{\end{eqnarray}}
\def\bs{\begin{subequations}}
\def\es{\end{subequations}}
\def\g{\gamma}
\def\G{\Gamma}
\def\vp{\varphi}
\def\mpl{M_{\rm P}}
\def\ms{M_{\rm s}}
\def\ls{\ell_{\rm s}}
\def\lp{\ell_{\rm pl}}
\def\l{\lambda}
\def\gs{g_{\rm s}}
\def\d{\partial}
\def\co{{\cal O}}
\def\sp{\;\;\;,\;\;\;}
\def\spa{\;\;\;}
\def\r{\rho}
\def\dr{\dot r}
\def\dt{\dot\varphi}
\def\e{\epsilon}
\def\k{\kappa}
\def\m{\mu}
\def\n{\nu}
\def\om{\omega}
\def\tn{\tilde \nu}
\def\p{\phi}
\def\vp{\varphi}
\def\P{\Phi}
\def\r{\rho}
\def\s{\sigma}
\def\t{\tau}
\def\x{\chi}
\def\z{\zeta}
\def\a{\alpha}
\def\b{\beta}
\def\de{\delta}
\def\bra#1{\left\langle #1\right|}
\def\ket#1{\left| #1\right\rangle}
\newcommand{\stt}{\small\tt}
\renewcommand{\theequation}{\arabic{section}.\arabic{equation}}
\newcommand{\eq}[1]{equation~(\ref{#1})}
\newcommand{\eqs}[2]{equations~(\ref{#1}) and~(\ref{#2})}
\newcommand{\eqto}[2]{equations~(\ref{#1}) to~(\ref{#2})}
\newcommand{\fig}[1]{Fig.~(\ref{#1})}
\newcommand{\figs}[2]{Figs.~(\ref{#1}) and~(\ref{#2})}
\newcommand{\GeV}{\mbox{GeV}}
\def\ricci{R_{\m\n} R^{\m\n}}
\def\riemann{R_{\m\n\l\s} R^{\m\n\l\s}}
\def\triemann{\tilde R_{\m\n\l\s} \tilde R^{\m\n\l\s}}
\def\tricci{\tilde R_{\m\n} \tilde R^{\m\n}}
\title{Single Field Baryogenesis and the Scale of Inflation}
\author{
K.R.S. Balaji $^{1)}$ \email[Email:]{balaji@hep.physics.mcgill.ca},
Robert H. Brandenberger $^{1,2)}$ \email[Email:]{rhb@hep.physics.mcgill.ca}
and Alessio Notari $^{1)}$ \email[Email:]{notari@hep.physics.mcgill.ca}}
\affiliation{1) Department of Physics, McGill University, Montr\'eal, QC, 
H3A 2T8, CANADA}
\affiliation{2) Physics Department, Brown University, Providence, RI 02912, USA}
\begin{abstract}
In the context of inflationary cosmology, we discuss a
minimal baryogenesis scenario in which the resulting baryon to entropy
ratio is determined by the amplitude of the anisotropies of the cosmic microwave
background. The model involves a new $SU(2)_L$ scalar field which
generates a Dirac neutrino mass and which is 
excited by quantum fluctuations during inflation, yielding a CP-violating phase.
During the scalar field decay after inflation, an asymmetry in the left-handed
neutrino number is generated, which then converts to a net baryon asymmetry via
sphalerons. A lower limit on the expected initial value of the scalar field
translates to a lower limit on the baryon to entropy ratio (which also depends 
on the Dirac neutrino Yukawa coupling). 
Consistency with the limits on baryonic isocurvature perturbations requires 
that the spectral index of 
adiabatic perturbations produced during inflation be very close to unity. In a variant
of our scenario in which the scalar 
is a gauge singlet, the connection between the baryon to entropy ratio and 
the inflationary scale is lost, although the basic mechanism of baryogenesis 
remains applicable.

\end{abstract}
\maketitle

\section{Introduction}

Two important numbers are measured in modern cosmology: the net baryon to entropy ratio
$n_B / s  \, \sim \, 10^{-10}$ (where $n_B$ is the difference between the number density
of baryons and antibaryons and $s$ is the entropy density) and the amplitude of the
angular power spectrum of the cosmic microwave background (CMB) temperature 
anisotropies,  $\delta T / T \, \sim \, 10^{-5}$. These two observations 
{\em independently require} new physics input from both
the particle physics sector and the pure cosmological sector. New particle physics 
input is required to explain the origin of the baryon asymmetry (this new
particle physics must satisfy the three Sakharov conditions \cite{sakharov}). The 
CMB anisotropies could be a natural consequence of inflation \cite{guth}. Again,
new particle physics appears to be required in order to generate inflation. 

In the initial models of baryogenesis (see e.g. \cite{Dolgov,Trodden} for reviews of
baryogenesis), no connection was made between baryogenesis and inflation.
A newer scenario for baryogenesis that establishes a 
connection with inflationary cosmology is the Affleck-Dine (AD) mechanism \cite{AD} 
or variants thereof. In this scenario, the required baryon asymmetry 
is generated via explicitly CP-violating and baryon number-violating 
decays of a new scalar field $\phi$, a field which may be related to the
field driving cosmological inflation.
However, no link between the value of $n_B / s$ and the amplitude of $\delta T / T$
was made.

Here, we study a simple variant of AD baryogenesis (proposed in \cite{BB}) which has
the special features that no new CP-violating phases and baryon number violating
interactions need to be introduced in the Lagrangian. Assuming that the observed
cosmological fluctuations were generated during a period of inflation, we find that
in our baryogenesis scenario the value of $n_B / s$ is determined by the amplitude of 
$\delta T / T$ (modulo a Yukawa coupling constant). Given a reasonable lower bound
on the initial value of the rolling scalar field $\phi$ yielding baryogenesis, we deduce a 
{\em lower bound} for the baryon asymmetry given the observed amplitude of the CMB
anisotropies. For reasonable values of the Yukawa coupling constant, the lower bound
yields the correct order of magnitude for the baryon to entropy ratio. 
The connection between the two observational values holds if $\phi$ is a 
$SU(2)_L$ doublet. We also examine the case where $\phi$ is a gauge singlet, 
and in this case the relation between the two observational quantities is lost
\footnote{There has been another recent interesting proposal \cite{Stephon}
in which the amplitudes of the baryon to entropy ratio and the CMB temperature 
fluctuations are related. This mechanism makes use of a helicity asymmetry in
the gravitational wave spectrum and is unrelated to ours.}.
 
\section{General considerations}

Our approach to baryogenesis relies on a simple variant of the 
AD mechanism that was recently introduced in \cite{BB}. 
Central to this scenario is a new complex scalar field $\phi$ which
obtains a non-vanishing expectation value as a consequence of
quantum fluctuations during inflation. 
If this field has motion in the phase direction, it acquires a charge which leads
to CP-violation and resulting leptogenesis. 
Note that $\phi$ neither carries non-vanishing baryon or lepton
charges, nor does it involve $CP$-violating couplings in the Lagrangian
\footnote{For earlier discussions on how to obtain cosmological CP-violation without
explicit CP-violating phases in the Lagrangian see \cite{Dolgov,BBBL}.}. 
We assume that the mass $m_{\phi}$ of $\phi$ is smaller than the Hubble expansion
rate $H_I$ during inflation. After inflation ends and the Hubble rate $H(t)$ has
fallen to $m_{\phi}$ the field begins to roll towards the minimum of its potential
$V(\phi)$. 

The charge asymmetry of $\phi$ which leads to leptogenesis is (see e.g. \cite{Trodden})
\footnote{Note that in the phase when $\phi$ is oscillating homogeneously in space,
the charge density $Im(\phi^{*} {\dot \phi})$ turns out to be of the same order of the number density of 
the $k = 0$ quanta of $\phi$ which the field configuration corresponds to. During a phase 
of slow rolling,
there is an extra factor which relates the number density to $Im(\phi^{*} {\dot \phi})$.
}
\be
Q_\phi \, =  \, Im(\phi^{*} \dot{\phi})\, .
\label{chrasym}
\ee
This charge vanishes if the potential is a function of $\phi^* \phi$. Thus, it is
essential for our mechanism that it has a different form, for example if it is a 
function of $\phi^2$ (plus complex conjugates). We consider two possibilities,
first the case when $\phi$ is a $SU(2)_L$ doublet, second when it is a singlet
under this gauge group. In the first case (the case studied in \cite{BB}
and which will be the main focus here), the potential leads to a violation of 
gauge invariance. 

Let us first consider the case when $\phi$ is a $SU(2)_L$ doublet. The key idea
is to view the gauge symmetry-violating potential as the result of a symmetry
breaking phase transition at a higher scale. Specifically, 
let us assume that at some large scale (say $\Lambda$) there is an extra
complex scalar field $\chi$ which is charged under $SU(2)_L$. The interaction
Lagrangian is taken to be
\be \label{intlag}
{\cal L} = y_{\nu} \bar\nu_L^i \nu_R^i \phi + 
V(|\phi|^2,|\sigma|^2, \phi^\dagger \sigma) + h.c.~.
\ee
which is gauge-invariant. We assume that at some scale
${\tilde \Lambda} < \Lambda$ the field $\sigma$ takes on a symmetry-breaking
expectation value $\langle\sigma\rangle$, 
and that it is massive and hence decouples on scales $\ll {\tilde \Lambda}$. 
The resulting effective Lagrangian automatically 
breaks the original gauge symmetry spontaneously and contains terms like $\phi^2$ 
\footnote{We are assuming that the 
vacuum expectation value of $\sigma$ does not break the SM symmetries 
in any other term.}. The original potential 
$V(|\phi|^2,|\sigma|^2, \phi^\dagger \sigma)$ reduces to a low energy effective
potential $V(|\phi|^2,|<\sigma>|^2, \langle \sigma\rangle \phi^\dagger)$. 
In this limit, it follows immediately that the rolling of $\phi$ leads to 
$Q_\phi \neq 0$. The case when $\phi$ is a singlet under $SU(2)_L$ will be
discussed in a separate section since the Yukawa interaction term in (\ref{intlag})
no longer can be used.

The Yukawa coupling in (\ref{intlag}) generates a Dirac mass for the neutrinos.
In a process analogous to what happens in models of leptogenesis, the Yukawa coupling
leads to the decay of $\phi$, in our case into Dirac 
neutrinos, producing an asymmetry in the left-handed leptons. This asymmetry is then
converted to the required baryon asymmetry via electroweak sphalerons 
transitions \cite{RubShap}. Note that in this mechanism, the total lepton number is not 
violated and the asymmetry in left-handed neutrinos is compensated by a corresponding 
asymmetry in the right-handed neutrino sector.

In the following we consider the usual particle physics standard model (SM) with the
addition of right handed neutrinos, and assume 
that the neutrinos, like any other fermion of the SM, obtain Dirac masses. 
The neutrino Yukawa coupling is small enough to prevent 
an equilibration between the left-handed and right-handed neutrinos (see \cite{BB}
for a discussion).
Note that one could also add a Majorana mass term for the right-handed neutrino $\nu_R$ 
without spoiling the mechanism.

In the first class of models, when $\phi$ is a $SU(2)_L$ doublet,
we find is a direct correlation between the produced baryon to entropy ratio and 
the amplitude of density perturbations generated during inflation. Indeed, 
we are able to provide a lower bound on $n_B / s$ which is in quantitative 
agreement with observations. 

On the other hand, in the second class of models, where $\phi$ is a singlet, 
the issue of gauge invariance is not a problem. The leading order interaction
terms are dimension five
non-renormalizable operators which describe the decay of $\phi$. 
If we allow for all possible interaction terms, it turns out that a 
sufficient baryon asymmetry can still be produced, although no longer with a
direct connection with the inflationary scale. 

\section{Connection between $n_B$ and $H_I$}

To establish the connection between the generated baryon asymmetry and 
the scale of inflation, we first estimate the baryon to entropy ratio produced 
from the initial charge asymmetry $Q_\phi$ (given in (\ref{chrasym}).
The field $\phi$ decays when $\Gamma=H$ (where $\Gamma$ is the total decay width), 
producing an asymmetry in the left-handed neutrinos (and  similar asymmetry with opposite
sign in the right-handed sector). The amount of asymmetry transferred to neutrinos is 
given by the charge evaluated at the decay time folded with the branching ratio, i.e.
\be \label{asym1}
\frac{(n_{\nu}-n_{  \bar{\nu}})}{s}= \frac{Q_{\phi}}{s} \frac{\Gamma_{\nu}}{\Gamma} ~,
\ee
where $s$ is the entropy density at the time $\Gamma\simeq H$, and $\Gamma_{\nu}$ is
the decay rate into neutrinos.
Generically, $m_\phi \gtrsim \Gamma$. Thus, 
at the time $H\simeq m_{\phi}$ when $\phi$ can begin to roll, it will
immediately start oscillating as 
matter around its minimum, and  its energy density gets diluted as $a^{-3}$
(where $a(t)$ is the cosmological scale factor) until 
$H\simeq \Gamma$. Now, evaluating (\ref{asym1}) at the decay temperature 
$T\simeq \sqrt{M_{pl} \Gamma}$ we obtain
\be \label{asym2}
\frac{n_B}{s}\approx Im[\phi^* \dot{\phi}] \frac{\Gamma_{\nu} }
{\Gamma^{5/2}M_{pl}^{3/2}}  \, .
\ee

{F}or the decay rate of $\phi$ into neutrinos, we take the usual perturbative result
\be \label{perturb}
\Gamma_{\nu}\approx y_{\nu}^2 m_\phi/(8 \pi)\, , 
\ee
and for the total decay rate we assume
\be
\Gamma \approx m_{\phi}^3/\phi^2 \, .
\ee
The latter result can be obtained by taking the usual perturbative decay rate as
in (\ref{perturb}) with an initially arbitrary coupling constant $g$, 
and taking for $g$ the maximal value for which the decay is kinematically allowed,
assuming that the mass of the decay product obtains a contribution $g <\phi>$
from the $\phi$ field. 
Note that this upper limit coincides with the decay
width used in the original AD model \cite{AD}.  In that case
 the only possible decay was to a light particle which has no direct coupling to 
 $\phi$ of the lighter final state, mediated via an intermediate heavy 
virtual state (to which a direct decay is forbidden) thereby, leading to a decay rate of 
the order (${\cal O}(m_{\phi}^3/\phi^2)$). However, a direct decay is also still possible 
if the coupling constants for the relevant interactions are sufficiently small.

With the results of the previous paragraph to obtain the branching fraction, 
we can evaluate the baryon to entropy ratio given by (\ref{asym2}). Since the field
rolling takes place at small field values, the potential can be approximated by
$V(\phi)\approx m_\phi ^2 \phi^2+h.c.$.  
Note that $\Gamma$ depends on time since the amplitude of $\phi$ depends on time
due to the red-shifting - $\phi \propto a^{-3/2} \propto H^{3/4}$ which means that at 
the decay time 
$\phi=\phi_0 \left(\Gamma /m_\phi \right)^{3/4}$ where, $\phi_0$ is the initial field 
value when $H \gtrsim m_\phi$. The condition $\Gamma=H$ is obtained for 
$H=\Gamma=m_\phi (m_\phi/\phi_0)^{4/5}$. Note that the energy density of the field 
$\phi$ is always sub-dominant as long as $m_\phi^2 \phi^2\gtrsim T^4$ or equivalently
$H\gtrsim m_\phi \left(\phi_0 /M_{pl}\right)^{4}$. This situation is allowed as long 
as the field values are somewhat smaller than $M_{pl}$ which is the limit of 
interest to us. If the field is light ($m_{\phi}<H_I$), its quantum fluctuations are
determined by $H_I$. To a first approximation, in a particular Hubble volume, $\phi$
undergoes a random walk with time step $H^{-1}$ and amplitude $H$ (see e.g.
\cite{Tsujikawa} for a recent discussion). The inflationary expansion effectively
homogenizes the value of $\phi$ in our patch of the universe. Note that in our case 
(for the quadratic potential), the real and imaginary components of the field
are uncorrelated and hence, it is likely that both these components will independently
take a value of order $H_I$. This leads to the relative phase to be of order unity. 
Therefore, in the oscillatory phase, the charge asymmetry is $ \sim m|\phi_0|^2$ 
multiplied by the dilution factor.

As a lower bound on the value of $\phi$ in our Hubble patch we take 
$\phi_0 \, \gtrsim \, H_I$. This leads to a lower bound on the value of $n_B / s$: 
\be
\frac{n_B}{s} \, = \, y_{\nu}^2 \left(\frac{\phi_0}{M_{pl}} \right)^{3/2}  
\left(\frac{\phi_0} {m_{\phi}} \right)^{13/10} \,
 \gtrsim  \, 10^{-9} y_{\nu}^2 \, ,
\ee
where we have made use of the observed amplitude of the CMB fluctuations to fix $H_I$.
We conclude that in the case when $\phi$ is a doublet case, a 
realistic value for $n_B / s$ comes directly from the amplitude of quantum fluctuations 
of $\phi$ during inflation. This is the main result of our analysis. 

%
\section{Constraints from isocurvature perturbations}

We examine here possible constraints on our baryogenesis scenario from the
current limits on the amplitude of the isocurvature perturbations which the above 
scenario would produce.

It is well known that perturbations in two different fields (namely an inflaton 
$\chi$ field and our charged field $\phi$) generically lead to isocurvature 
perturbations if the two fields decay into different types of particles.
This may lead to observable consequences in the measurements of  CMB anisotropies and  
Large Scale Structure. Present experiments are consistent with a purely adiabatic spectrum, 
and already put constraints on the amount of isocurvature perturbations which may be 
present. Moreover, it is expected that future experiments may be able to detect even
small contributions of isocurvature perturbations, providing a valuable tool for the 
investigation of the early Universe.

To study the fluctuations, we expand the metric to linear order about a FRW background.
In longitudinal gauge, the line element is (see \cite{pert} for a review)
\be \label{pertmetric} 
ds^2 \, = \, (1 + 2 \psi)dt^2 - a^2(t)(1 - 2 \psi)\delta_{ij}dx^i dx^j \, ,
\ee
where $\psi$ is a function of space and time and contains the information about scalar
metric fluctuations (we are assuming the absence of anisotropic stress).
The curvature perturbation $\zeta$ on a comoving
hyper-surface (which measures the adiabatic fluctuation) is given by \cite{pert}
\be
\zeta \, \equiv \, \psi + H\frac{\delta\rho}{\dot{\rho}} \, 
\label{zeta}
\ee
(Note that this quantity is independent of gauge).
Here, $\delta\rho$ is the perturbation of the total energy density. On super-Hubble scales 
and in the absence of isocurvature perturbations, $\zeta$ is a conserved quantity.
Inflationary cosmology predicts an approximately scale-invariant spectrum of 
fluctuations. Thus, neglecting the small deviations from scale-invariance, the Fourier
modes of $\zeta$ are given by
\begin{equation}
\zeta(k) =  \frac{\sqrt{4 \pi}}{ \sqrt{\epsilon_{tot}} M_{pl}} A_k e_A(k)~,  
\label{usualzeta}
\end{equation}
where $A_k \equiv {H}/{\sqrt{2 k^3}}$ (using the convention that comoving wavenumbers
are dimensionless), and $e_R(k)$ is a Gaussian random variable 
normalized to one \cite{gordonwands}. Here,
\be
\epsilon_{tot}\equiv \epsilon_{\chi} + \epsilon_{\phi} \, {\rm and} \,
\epsilon_I\equiv \frac{M_{pl}^2}{16 \pi} 
\left(\frac{\partial V/\partial \phi_I}{V}\right)^2 
\ee
is the slow roll parameter for the field $\phi_I$ (note that because of the
small value of the mass $m_{\phi}$, $\phi$ will be slowly rolling during inflation). 
In the above, $V(\phi,\chi)$ is the potential for 
the fields during inflation.
when the presence of the field $\phi$ during inflation 
does not change this prediction (this is true for example if both fields roll with 
constant velocity during inflation \cite{gordonwands}).
This means that the inflationary adiabatic perturbation is not seeded  by isocurvature 
perturbations and thus they are statistically uncorrelated.

Moreover, in our scenario, the energy density of the field $\phi$ is always sub-dominant 
with respect to the radiation which is produced by inflaton decay.
As a consequence, after inflation, the perturbations in the field $\phi$, and thus also 
the fluctuations in 
the energy density of its decay products, do not contribute appreciably to 
$\zeta$ \cite{curvaton}. This means that we use as value for $\zeta$ at late times, the 
usual expression as in (\ref{usualzeta}).

Following the above arguments, the interesting quantity for us is the baryon to photon 
isocurvature perturbation
\begin{equation}
S_{B \gamma} \, \equiv \, \frac{\delta(n_B/s)}{n_B/s}
\, = \, \frac{\delta\rho_B}{\rho_B}-\frac{3}{4}
\frac{\delta\rho_{\gamma}}{\rho_{\gamma}} ~,
\label{sbg}
\end{equation}
which is by construction a gauge invariant quantity. Alternatively, we can express 
(\ref{sbg}) as \cite{curvaton}
\begin{equation}
S_{B \gamma}=3(\zeta_B-\zeta_{\gamma})
~,
\end{equation}
where 
$
\zeta_i\equiv\psi-H\frac{\delta \rho_i}{\dot{\rho}_i}~.
$
Now, since $\zeta_{\gamma}$ is a conserved quantity, it is
equal to its initial value $\zeta_{\chi}$ which is given at the end of inflation. For 
baryons, following \cite{curvaton}, we introduce the quantity
$
\tilde{\zeta}_B \, \equiv \, \psi-H\frac{\delta n_B}{\dot{n}_B}~.
$
For non-relativistic particles we have $\zeta_{B}=\tilde{\zeta}_B$. This holds for 
baryons after the QCD phase transition.
Once the field $\phi$ decays, a non vanishing $\tilde{\zeta}_B$ is created
leading to a relation 
$
\tilde{\zeta}_B \, \simeq \, \zeta_{\phi} \, .
$
Moreover, at all times before the decay $\zeta_{\phi}$ is a conserved quantity, as long 
as there is no energy transfer between $\phi$ and other fluids.
This means that $S_{B \gamma}$ is directly related to $\zeta_{\phi}$ and $\zeta_{\chi}$ 
whose values are fixed by the inflationary phase. In particular \cite{gordonwands,bartolo}
\begin{equation}
|S_{B \gamma}| \, \simeq \, | S_{\phi \chi}| \, \simeq  \, 3 \frac{\sqrt{4 \pi}} {M_{pl}} 
 \frac {\sqrt{\epsilon_{tot}}}  {\sqrt{\epsilon_{\phi}} \sqrt{ \epsilon_{\chi} }  }     
A_k |e_S(k)| ~,
\end{equation}
where $e_S$ is a normalized Gaussian random variable.
The quantity which is constrained by experiments is the ratio of the amplitudes 
of $S_{B \gamma}$ and $\zeta$, which has to be less than about three if the  
isocurvature and adiabatic perturbations are uncorrelated \cite{garcia}. 
This translates into the constraint
\begin{equation}
\left\arrowvert {{S_{B \gamma}} \over {\zeta}} \right\arrowvert \, \simeq \,
\frac{3\epsilon_{tot}}{\sqrt{\epsilon_{\chi}     \epsilon_{\phi}  }} \lesssim 3~. 
\label{slow}
\end{equation}
In fact, evaluating the slow-roll parameter $\epsilon_{\phi}$ for our quadratic potential 
and using the CMB constraint on the overall amplitude of scalar perturbations 
\cite{Smoot:1992td} we obtain
\begin{equation}
\left\arrowvert {{S_{B \gamma}} \over {\zeta}}\right\arrowvert \, \simeq \, 
6 \sqrt{\pi} \frac{(7\cdot 10^{16} GeV)^4}{m_{\phi}^2 \phi_0 M_{pl}}  
\frac{\epsilon_{tot}^2} {\sqrt{\epsilon_{\chi} } }  \lesssim 3~.  \label{isocurv}
\end{equation}
This implies that our scenario is viable only if $\epsilon_{tot}$ during inflation is a 
very small number, which in turn implies that the spectral index has to be very 
close to one. 
%
\section{Unbroken gauge symmetry}

Finally, let us present a model where the complex scalar $\phi$ is an $SU(2)_L$ singlet. 
In order to couple $\phi$ with neutrinos, one has to include higher dimension terms, for 
instance,
\begin{equation} 
{\cal L}_{\phi H \nu} \, = \, \frac{\lambda_{ij}}{\Lambda} H \bar\nu_L^i \nu_R^j \phi 
+h.c.~. 
\end{equation}
Here, $\Lambda$ is some cutoff scale, and $\lambda_{ij}$ are couplings of order unity.
In this scenario, the four-particle interaction generates the lepton asymmetry, through 
the decay of the field $\phi$. Let us now analyze the number $n_B/s$ in this case.
The decay width is suppressed by the five dimensional interaction term. Hence,
in order to obtain the observed baryon to entropy ratio, we  
require a value of $\phi_0$ which is larger than $H_I$. Since $\phi$ undergoes a random
walk with amplitude $H_I$ during inflation, it is easy to obtain larger values of
$\phi_0$ provided $m_{\phi}$ is substantially smaller than $H_I$ (see e.g. \cite{Tsujikawa}
for a recent discussion).

Given a value for $\phi_0$, and making use of an estimate for the asymmetry of the
decay rate into neutrinos during the process $\phi\rightarrow H \nu_L \nu_R$ which
we take to be $\Gamma_{\nu}\approx m_{\phi}^3/\Lambda^2$, we obtain 
(following the same logic as in the third section)
\be
\frac{n_B}{s} \approx  \frac{\phi_0^{14/5} m_{\phi}^{7/10}}
{M_{pl}^{3/2} \Lambda^2}  ~.
\label{singletnB}
\ee
Assuming $\Lambda\approx M_{pl}$, and for sufficiently large field values and masses, 
(\ref{singletnB}) can yield the correct value for $n_B/s$.
If $m_{\phi}$ is smaller than $H_I$, we expect isocurvature perturbations 
to be again given by eq.~(\ref{slow}) 
(as before, a small slow roll parameter is required to be 
consistent with observations).

\section{Discussion and Conclusions}

To summarize, in the minimal model of baryogenesis proposed in \cite{BB},
in which a rolling scalar field obtains a CP-asymmetry due to its initial
conditions, and in which the decay of this field generates an asymmetry
in the left-handed neutrino number density, 
we have shown that the baryon to entropy ratio is determined by 
the amplitude of CMB temperature fluctuations, thus connecting two
observational numbers from cosmology which usually are considered not to be
related. This relation holds if the scalar
field is charged under $SU(2)_L$. More specifically, making use of the observed
amplitude of the CMB anisotropies we obtain a lower bound on
the baryon to entropy ratio which agrees quite well with the observed value. The
exact value also depends on the magnitude of the neutrino Yukawa coupling. 

We have also analyzed the constraints on our model coming from the observational
limits on isocurvature perturbations. We find consistency provided that  
the adiabatic perturbations have a spectral index very close to the scale invariant one.
Moreover, evidence for baryon isocurvature perturbations
may be a signal for this model, giving information on the predicted baryon asymmetry. 
In order to completely calculate the baryon asymmetry, the knowledge of the Yukawa coupling 
of the new $SU(2)_L$ scalar with neutrinos is required. The link between $n_B / s$ and
$\delta T / T$ is lost if we consider the scalar field driving baryogenesis to be a
gauge singlet (which we illustrated using a dimension five interaction). 

{\bf Acknowledgements}:

This work is supported by NSERC (Canada) and by the Fonds 
de Recherche sur la Nature et les Technologies du Qu\'ebec.
The work of RB has also been supported in part (at Brown) by the 
US Department of Energy under Contract DE-FG02-91ER40688, TASK A.

\end{document}